\documentclass[a4paper,12pt]{article}

\usepackage[T2A]{fontenc}
\usepackage{amsmath,amssymb}
\usepackage{amsfonts}
\usepackage{euscript}
\usepackage{cite}
\usepackage[unicode,linktocpage=true,plainpages=false,pdfpagelabels=false]{hyperref}

\DeclareMathOperator{\tr}{Tr}
\DeclareMathOperator{\sgn}{sgn}
\renewcommand{\Im}{\mathop{\mathrm{Im}}\nolimits}
\renewcommand{\Re}{\mathop{\mathrm{Re}}\nolimits}

\begin{document}

\title{The foundations of quantum theory\\ and its possible generalizations\footnote{
The paper was written based on the materials of the lecture on the 15th International V.A. Fock School for Advances of Physics 2005, 21-27 November 2005, St. Petersburg.
Published in the Proceedings of the School, ed. by V. Novozhilov, Publishing house of Saint Petersburg State University, St.~Petersburg, 2006.
}}

\author{V.~A.~Franke\thanks{E-mail: valentin.alf.franke@gmail.com}\\
{\it Saint Petersburg State University, Saint Petersburg, Russia}
}
\date{\vskip 15mm}
\maketitle

\begin{abstract}
Possible generalizations of quantum theory permitting to describe
in a unique way the development of the quantum system and the
measurement process are discussed. The approach to the problem
based on the Lindblad's equation for the statistical operator is
reviewed. The Tomonaga-Schwinger like equation of this type is
introduced to establish Lorentz invariance. The application of
tachyonic field to overcome divergences arising in this equation
is analyzed. Other approaches to the problem are shortly
discussed.
\end{abstract}

\newpage

Since the discovery of quantum mechanics Albert Einstein
and some other physicists were not satisfied with its standard
(Copenhagen) interpretation.
The interest in this field rises today because (a) the theorists
have approached in their constructions the Planck scale ($l_{pl}
\approx 10^{-33}$ cm; $m_{pl} \approx 10^{-5}$ gr), where a new
physics may be found; (b) the astronomical observations carried
out by man-made satellites permit to verify cosmological models
describing the very beginning of the Universe history, when the
observable part of the Universe was much smaller than an atom;
hence, to
construct such models one has to apply quantum theory to the
Universe as a whole; this requires a generalization of conventional
interpretation of quantum theory; (c) intensive
investigations of the domain, intermediate between microscopic and
macroscopic world, are now carried out; this is related to the
interpretation of quantum theory.

In the period of eighty years some physicists have formulated the
following requirements which are desirable to be
fulfilled and which the quantum theory does not satisfy
or does not exactly satisfy: (1) the theory should describe in a unified way
the existing reality and
not only the relations between observations
(as J.~S.~Bell said, ``the theory should be not about observables
but about beables''); (2) being formulated in terms of
probabilities the quantum theory should permit a statistical
derivation; an unknown today subquantum world should be found, where the
corresponding statistics takes place; (3) some conditions of
locality and causality should be fulfilled in the mentioned
subquantum world \cite{1}.

We shall concentrate ourselves mainly on the first point because
it is a necessary prerequisite to the investigations of the second
and third points and because it is very little known about the
second point. But beforehand we shall describe the important
result of J.~S.~Bell concerning causality in the hypothetical
subquantum world.

Let two spin $\frac12$ particles are prepared in a common
state of spin 0 in the spacetime region $\sigma$ at the time
$t_0$. Let the particles move to the regions A and
B correspondingly, where two observers measure simultaneously at
the time $t_1 > t_0$ the spin projections of these particles onto the
directions $\vec A$ and $\vec B$. Let the spin-orbital interaction
be absent. Then the quantum mechanics teaches us and the
experiment confirms that the probability to find both spin
projections to be positive is equal to $w(\vec A, \vec B) = \sin^2
\frac{\theta}{2}$, where $\theta$ is the angle between the
directions $\vec A$ and $\vec B$.

Trying to explain in an objective way the correlation between
the measurements in the regions A and B one introduces hidden
variables existing in the region $\sigma$. Let us denote all these
variables by one symbol $\alpha$. Then one assumes that there exists a
probability density $\rho(\alpha)$ such that

\begin {equation} \label{density}
\rho(\alpha) \ge 0, \qquad \int d\alpha \rho(\alpha) = 1.
\end {equation}

Furthermore, one assumes that for a definite value $\alpha$ of
hidden variables the  conditional
probabilities $P(\alpha, \vec A)$ and
$P(\alpha, \vec B)$ exist to get positive results when the
spin projections in the directions $\vec A$ and $\vec B$ are
measured. These probabilities fulfill the conditions

\begin {equation} \label{conditionsA}
P(\alpha, \vec A) \ge 0, \qquad P(\alpha, \vec A) + P(\alpha, - \vec A) = 1,
\end {equation}
\begin {equation} \label{conditionsB}
P(\alpha, \vec B) \ge 0, \qquad P(\alpha, \vec B) + P(\alpha, - \vec B) = 1,
\end {equation}
for all $\vec A$ and $\vec B$. From the conventional probability
theory it follows that

\begin {equation} \label{bothpositive}
w(\vec A, \vec B) = \int d\alpha \rho(\alpha) P(\alpha, \vec A) P(\alpha, \vec B).
\end {equation}

J.~S.~Bell has shown that no functions exist fulfilling the
conditions \eqref{density}, \eqref{conditionsA},
\eqref{conditionsB} and the equality $w(\vec A, \vec B) = \sin^2
\frac{\theta}{2}$ even approximately (the discrepancy turns out to be
about 30\%). Here it is important that $P(\alpha, \vec A)$ does
not depend on $\vec B$ and $P(\alpha, \vec B)$ does not depend on
$\vec A$ due to relativistic causality which one assumes to
hold in the subquantum world of hidden variables.

It follows that no theory of subquantum world where the
conventional relativistic causality holds can be built. In every
theory of subquantum world the relativistic causality must arise
at higher level of quantum and macroscopic world.

Returning to the problem of unified description of reality it is
natural to assume the following principle of macroscopic
definiteness: there exists really only one four-dimensional
macroscopic picture of the world where all macroscopic quantities
are defined with macroscopic precision.

The most straightforward way to implement this principle and to
ensure simultaneously the unified description of reality is related
 with the generalization of quantum theory.
One has to change the equation describing the time evolution of
 quantum system in such a way that each superposition of
macroscopically different quantum states transforms in time automatically
into one of these states with conventional probabilities. The
description of microscopic systems should not change
significantly. The generalized theory should be irreversible in
time. So one has to modify the Schr\"odinger equation which is
reversible. The simple possibility is to use the equation for
statistical operator (quantum density matrix) $\rho$. This permits to
describe irreversibility. We consider firstly such a possibility in
formal manner and afterwards discuss some physical problems and
other approaches.

If it is known that an object possesses a quantum state
vector (a wave function) $\psi_n$ with
the probability $w_n$, we attribute to it a $\rho$-matrix

\begin {equation} \label{rhomatrix}
\rho = \sum_n w_n |\psi_n\rangle\langle\psi_n|.
\end {equation}
When $\rho = |\psi\rangle\langle\psi|$ one calls the state of the
object pure and otherwise mixed. The average value of any hermitian
operator $A$ in the state $\rho$ is equal to

\[
tr(A\rho) = \sum_n w_n \langle\psi_n|\rho|\psi_n\rangle.
\]
The representation \eqref{rhomatrix} of a mixed state is not
unique. It becomes unique if one requires that
$\langle\psi_n|\psi_m\rangle = \delta_{nm}$ and if all the $w_n$
are different. The $\rho$-matrix characterizes the mixed state
completely. No more information can be extracted from such state
by experiment than it is contained in the $\rho$-matrix.

In the conventional quantum theory the $\rho$-matrix fulfills the
equation

\begin {equation} \label{rhomatrixSchr}
\frac{d\rho}{dt} = -i[H, \rho],
\end {equation}
which is equivalent to  Schr\"odinger equation for the wave
functions $\psi_n$. $[,]$ is a commutator. The $\rho$-matrix is
hermitian ($\rho = \rho^{\dagger}$), normalized ($\tr\rho = 1$)
and positive definite ($\langle\psi|\rho|\psi\rangle \ge 0\;\; \forall\; \psi$).
The equation \eqref{rhomatrixSchr} preserves
these properties and is reversible in time.

The solution of the equation \eqref{rhomatrixSchr} has the form
\begin{equation} \label{evolution}
\rho(t) = U(t)\rho(0)U^{\dagger}(t),
\end {equation}
where the unitary operator $U(t)$ is equal to $U(t) = \exp(-iHt)$.
The transformation \eqref{evolution} is linear and preserves the
hermicity, normalization and positive definiteness of $\rho$. It
describes a process reversible in time.

To introduce the irreversibility in time one has to generalize the
transformation \eqref{evolution}. It must stay linear to conserve
the probabilities $w_n$ in the eq-n \eqref{rhomatrix}. These
probabilities describe our knowledge of the system and cannot
change in time. Furthermore the hermicity, normalization and
positive definiteness of $\rho$ should be preserved as before to
avoid complex and negative probabilities and to conserve the
probability.

It is known for a long time \cite{2} that a most general linear
transformation of the $\rho$-matrix preserving the hermicity and
normalization can be written in the form

\begin {equation} \label{mostgeneral}
\rho(t) = \sum_n\lambda_nA_n\rho(0)A_n^{\dagger},
\end {equation}
where $\lambda_n$ are real numbers and $A_n$ are operators such
that $\sum_n\lambda_nA_n^{\dagger}A_n = I$.

If one takes
\begin {equation} \label{positivelambda}
\lambda_n > 0 \quad\forall\; n,
\end {equation}
the positive definiteness is also preserved, because it
follows from the equality

\[
\rho(0) = \sum_m w_m |\psi_m\rangle\langle\psi_m|, \qquad w_m > 0,
\]
that in this case
\[
\langle\psi|\rho(t)|\psi\rangle = \sum_{n, m} w_m \lambda_n \langle\psi|A_n|\psi_m\rangle \langle\psi_m|A_n^{\dagger}|\psi\rangle \ge 0 \quad \forall\; \psi.
\]
But under the condition \eqref{positivelambda} the relation
\eqref{mostgeneral} is not the most general transformation
preserving the positivity condition. For example a transformation

\begin{equation} \label{specialexample}
\rho(t) = \sum_n \lambda_n A_n \rho(0)A_n^{\dagger} + \sum_m \tilde \lambda_m (B_m \rho(0) B_m^{\dagger})^*,
\end {equation}
where
\[
\sum_n \lambda_n A_n^{\dagger}A_n+\sum_m\tilde\lambda_m B_m^{\dagger}B_m=I,
\]
$\lambda_n > 0, \tilde\lambda_m > 0 \;\; \forall\; n,
m$ and $*$ designates complex conjugation, preserves positivity,
hermicity and normalization, but cannot in general be written in
the form \eqref{mostgeneral} with $\lambda_n > 0$.

One calls each linear transformation of $\rho$-matrices preserving
hermicity, normalization and positivity conditions a positive
dynamical transformation, and one calls the transformation of the
form \eqref{mostgeneral} with $\lambda_n > 0 \;\; \forall\;
n$ completely positive dynamical transformations.

The role of completely positive transformations is seen from the
following example\label{decoherence}. Consider a system consisting
of two objects which do not interact before the time moment
$t=0$ possess at this moment the $\rho$-matrices $\rho_{AB}^{(1)}(0)$
and $\rho_{ik}^{(2)}(0)$. Common $\rho$-matrix at $t=0$ is the
product

\[
\rho_{Ai, Bk}(0) = \rho_{AB}^{(1)}(0) \rho_{ik}^{(2)}(0).
\]
Let these objects interact in the time interval $(0, t_1)$ and let
the evolution of the system be described in this interval by a
unitary operator $U_{Ai, Bk}$. Then at the moment $t_1$ we have the
$\rho$-matrix

\[
\rho_{Dl, Em} = \sum_{A,i,B,k} U_{Dl, Ai} \rho_{AB}^{(1)}(0) \rho_{ik}^{(2)}(0) U_{Bk, Em}^{\dagger}.
\]
If we make observations at the time $t_1$ only on the first object
and are not interested in the second, we can get the $\rho$-matrix
of the first object putting

\[
\rho_{DE}^{(1)}(t_1) = \sum_l \rho_{Dl, El}(t_1) = \sum_{l,i,k}\sum_{A,B} U_{Dl, Ai} \rho_{AB}^{(1)}(0) \rho_{ik}^{(2)}(0) U_{Bk, El}^{\dagger}.
\]
Because
\[
\rho_{ik}^{(2)}(0) = \sum_p \lambda_p |\psi_p(0)\rangle_{i}\;{}_{k}\langle \psi(0)|, \qquad \lambda_p > 0,
\]
one gets
\begin{equation} \label{terriblemess}
\rho_{DE}^{(1)}(t_1) = \sum_{p,l} \lambda_p \left( \sum_i U_{Dl, Ai} |\psi_p(0)\rangle_i \right) \times \rho_{AB}^{(1)}(0) \left( \sum_k  {}_{k}\langle\psi_p | U_{Bk, El}^{\dagger} \right).
\end{equation}

This is a completely positive transformation of the $\rho$-matrix
$\rho^{(1)}$. It is irreversible even when the initial states
$\rho^{(1)}(0)$ and $\rho^{(2)}(0)$ are pure because of the
summation over the index $l$ in the eq-n \eqref{terriblemess}. Let us
stress that the irreversibility arises not because the two systems
interact but because the information about the second system is
lost completely. And this is a common rule: the development of a
state in quantum theory becomes irreversible if some information
about the considered system disappears.

The described situation is typical for irreversible physical
processes. That's why one assumes commonly that only completely
positive transformations are of physical interest. But this is not
exactly the case. In quantum field theory one has often to do
with indefinite metric. In this case the operators $U_{Dl, Ai}$
are only pseudounitary and the transformation \eqref{terriblemess}
may not be completely positive. Nevertheless under appropriate
conditions it can be positive. This is a special way to get rid
of the indefinite metric by introducing irreversibility. That's
why the positive but not completely positive dynamical
transformations are worth of some attention. But it is very hard
unsolved mathematical problem to find the general form of positive
dynamical transformations. Today it is known that the equation
\eqref{specialexample} defines a most general positive
transformation only for a $\rho$-matrices of dimension $2 \times
2$ \cite{3} but not in other cases. The problem of describing general
positive transformations is connected with Artin's theorem which
solves the 5-th Hilbert problem. Artin has proven that each
rational function which is positive everywhere can be represented
as a sum of squares of rational functions.

Let us write the dynamical transformation in the form
\[
\rho_{ik}(t) = \sum_{lm} L_{ik, lm} \rho_{lm}(0),
\]
where $L_{ik, lm} = L_{ki, ml}^*$ to ensure hermicity. It is
enough to preserve positivity for pure initial states $\rho_{lm} =
\psi_i\psi_m^*$. The positivity condition in this case looks like

\[
\sum_{i,k,l,m} {\psi_k'}^*\psi_i' L_{ik,lm} \psi_l \psi_m^* \ge 0 \quad \forall\; \psi_i', \psi_l.
\]
According to Artin's theorem there exists a representation
\begin{equation} \label{artin}
\sum_{i,k,l,m} {\psi_k'}^*\psi_i' L_{ik,lm} \psi_l \psi_m^* = \sum_{\nu}\left(\frac{P_{\nu}^*P_{\nu}}{Q_{\nu}^*Q_{\nu}}\right),
\end{equation}
where $P_{\nu}, Q_{\nu}$ are polynomials with respect to $\psi_i,
\psi_k', \psi_l^*, \psi_m^*$. Here the extension of Artin's
theorem to complex region is used. After the right side of the
equation \eqref{artin} is reduced to a common denominator this
denominator should cancel with the nominator. How to fulfill this
condition in general is unknown because the proof of Artin's
theorem is based on Zorn's lemma and that's why it is unconstructive.
From here on we consider only completely positive dynamical
transformations.

After the differentiation of the relation
\[
\rho(t) = \sum_n\lambda_n(t)A_n(t)\rho(0)A_n^{\dagger}
\]
with respect to time and introduction of some new denotations one
gets the equation

\begin{equation} \label{lindblad}
\frac{d\rho}{dt} = -i[H, \rho] + \sum_n \alpha_n \left( 2B_n\rho B_n^{\dagger} - B_n^{\dagger}B_n\rho - \rho B_n^{\dagger} B_n \right), \qquad \alpha_n > 0,
\end {equation}
where $B_n$ are some operators. It is easy to verify directly that
this equation preserves hermicity, normalization and positivity.
The equation \eqref{lindblad} is called the Lindblad's equation \cite{4}.

Via this equation one can describe the decay of the
macroscopically indefinite state into macroscopically definite
ones. Let us describe an example. Consider the simplest
case when only one operator $B$ is present in \eqref{lindblad} and
$H=0$. Then
\[
\frac{d\rho}{dt} = \alpha \left( 2B\rho B^{\dagger} - B^{\dagger}B\rho - \rho B^{\dagger} B \right).
\]
Let the operator $B$ be hermitian, so that $B=B^{\dagger}$.
Consider the frame where $B$ is diagonal. Let $b_1, b_2 \ldots$
are the eigenvalues of $B$. Then
\[
\frac{d (b_1|\rho|b_2)}{dt} = \alpha \left( 2b_1(b_1|\rho|b_2)b_2 - b_1^2(b_1|\rho|b_2) - (b_1|\rho|b_2)b_2^2 \right).
\]
or
\[
\frac{d (b_1|\rho|b_2)}{dt} = - \alpha (b_1 - b_2)^2(b_1|\rho|b_2).
\]
The solution is
\[
(b_1|\rho(t)|b_2) = \exp(-\alpha(b_1 - b_1)^2t) \times (b_1|\rho(0)|b_2).
\]
So all the nondiagonal matrix elements of $\rho$ disappear when
the time goes and the matrix $\rho$ becomes a mixture of
eigenstates of the operator $B$.

The constant $\alpha$ should be very small because otherwise the
conventional quantum mechanics of microscopic objects will be
destroyed. But when $B$ is a macroscopic operator, the quantity
$\alpha(b_1-b_2)^2$ may be large, and the initial state
may decay into eigenstates of this operator.

Now let us return to the general equation \eqref{lindblad}. Let
all the operators $B_n$ be macroscopical and all the numbers
$\alpha_n$ very small. The operators $B_n$ may not commute exactly
with the Hamiltonian $H$ and with each other. But the average
values of these commutators are much smaller than the average
values of the operators $B_n$. So one should expect that the
superposition of the eigenstates of operators $B_n$ belonging to
macroscopically different eigenvalues will be destroyed.

We see that via replacing the Schr\"odinger equation by the
equation \eqref{lindblad} with appropriate operators $B_n$ and
numbers $\alpha_n$ we can formally fulfill the principle of
macroscopic definiteness in the case of nonrelativistic physics.

The other proposed approach to the problem is based on a
stochastic differential equation for the quantum state vector
$\psi$ \cite{5}. One assumes that $\psi$ is a stochastic quantity and
writes for it an equation similar to Schr\"odinger's but with some
noise. This approach is equivalent to considering the probability
distribution $W(\psi)$ over the Hilbert space of all quantum state
vectors $\psi$.

Clearly
\begin{equation} \label{Wpositive}
W(\psi) \ge 0
\end{equation}
and
\begin{equation} \label{Wnorm}
\int d\mu(\psi)W(\psi) = 1,
\end{equation}
where $d\mu(\psi)$ is some measure on the Hilbert space. For this
distribution one may write an evolutionary equation

\begin{equation} \label{Wevolution}
\frac{dW(\psi)}{dt} = L(W(\psi))
\end{equation}
where $L$ is some linear operator preserving the conditions
\eqref{Wpositive}, \eqref{Wnorm}. This approach has the advantage
that one can directly restrict the functional $W(\psi)$ to be zero
on macroscopically indefinite states. But the following additional
restriction must be fulfilled. If two distributions $W_1(\psi)$
and $W_2(\psi)$ correspond to one and the same $\rho$-matrix at
the initial time, i.e.

\begin{equation*}
\begin{split}
\rho(t_1) &= \int d\mu(\psi) W_1(t_1, \psi) |\psi\rangle\langle\psi| \\
          &= \int d\mu(\psi) W_2(t_1, \psi) |\psi\rangle\langle\psi|,
\end{split}
\end{equation*}
then this equality should persist with time. To find the general
condition under which such requirement takes place is equivalent
to discover the general form of positive dynamical transformation.
It is extremely difficult.

That's why it seems easier to write firstly the equation
\eqref{lindblad} for the $\rho$-matrix and then to verify whether
it can be represented in the form \eqref{Wevolution} with
appropriate $W(\psi)$. If not, one has to look for other equation
of the type \eqref{lindblad}. Let us remark that one and the same
equation \eqref{lindblad} can be represented in the form
\eqref{Wevolution} in many ways, because the representation of the
$\rho$-matrix \eqref{density} is not unique.

One further approach, popular today, is called ``the method of
decoherent histories''. One assumes that the Universe behaves such
as if somebody measures periodically or continuously some set of
macroscopic quantities. Let these ``measurements'' take place at
the time moments $t_1, t_2, \ldots$ At each moment $t_i$ one
defines a set of projectors $p_{in}$ on subspaces of the Hilbert
space, such that

\begin{gather*}
p_{in}p_{im} = p_{im}p_{in} = 0 \quad \forall\; m \ne n, \forall\; i \\
\sum_n p_{in} = I \quad \forall\; i.
\end{gather*}
Each subspace corresponds to definite values of macroscopic
quantities fixed with macroscopic precision but is large enough to
contain the superpositions of microscopic states. One assumes that
the probability that the macroscopic quantities have corresponding
values is equal to

\[
\langle\psi|p_{1n_1}p_{2n_2}\ldots p_{N-1, n_{N-1}} p_{N, n_N} p_{N-1, n_{N-1}} \ldots p_{2n_2} p_{1n_1} | \psi \rangle,
\]
where $\psi$ is the initial state of the Universe. We have used
the Heisenberg representation. This approach gives for each
quantity $\langle \psi |p_{in}| \psi \rangle$ the exact
conventional quantum value only if all projectors $p_{in}$ commute
in the Heisenberg representation (for all $i$ and $n$). This is
very difficult to achieve because to do this one has to solve the
Heisenberg equations of motion. But one may hope that, if the
projectors $p_{in}$ are macroscopic and do not exactly commute,
the deviation of the values of $\langle \psi |p_{in}| \psi
\rangle$ from the conventional quantum values will be very small.

This approach is connected with the method based on the Lindblad's
equation. If one appropriately defines the $\rho$-matrix at each
moment $t_i$, introduces the Schr\"odinger representation and
goes to the limit $t_{i+1} - t_i \rightarrow 0$, one gets the
Lindblad's equation \eqref{lindblad}.

Let us now take into account the requirement of the Lorentz
invariance assuming that the space-time is flat. The simplest way
to do this consists in going to interaction representation
assuming that the field operators fulfill the conventional
equations for free fields. We consider the $\rho$-matrix
$\rho(\sigma)$ which depend on the spacelike hypersurface $\sigma$
and write down an equation similar to one of Tomonaga-Schwinger:

\begin{equation} \label{TS}
\begin{split}
\frac{\delta \rho(\sigma)}{\delta \sigma(x)} &= -i[H_{int}(x), \rho] + \\
&+ \alpha \sum_n \left\{2A_n(x)\rho(\sigma)A_n^{\dagger}(x) - A_n^{\dagger}(x)A_n(x)\rho(\sigma) - \rho(\sigma)A_n^{\dagger}(x)A_n(x) \right\}.
\end{split}
\end{equation}

To fulfill the Bloch integrability condition one has to assume
that all operators $H_{int}(x)$, $A_n(x)$, $A_n^{\dagger}(x)$
commute when they are taken at different points of the
hypersurface $\sigma$. This is not possible if $A_n$ are
macroscopical operators as we have assumed in the nonrelativistic
case. Nevertheless it is worth to investigate the case when the
$A_n$ are local operators commuting at different points on
$\sigma$.

For simplicity let us consider the case, when only one free
complex scalar field is present, and write the equation \eqref{TS}
as follows

\begin{equation} \label{TSsimple}
\frac{\partial \rho(\sigma)}{\partial \sigma (x)} = \alpha (2\phi(x)\rho(\sigma)\phi^{\dagger}(x) - \phi^{\dagger}(x)\phi(x)\rho - \rho\phi^{\dagger}(x)\phi(x)).
\end{equation}
Let $\rho$ be a vacuum state
\[
\rho = |\Omega\rangle\langle\Omega|
\]
Taking into account that
\[
\phi(x) = \int \frac{d^3p}{\sqrt{2p_0}}\bigl(b^{\dagger}(\vec p) e^{ipx} + a(\vec p) e^{-ipx}\bigr),
\]
where $b^{\dagger}$, $a$ are creation and annihilation operators,
one sees that the term $\phi(x)\rho\phi^{\dagger}(x)$ renders the
vacuum state $|\Omega\rangle\langle\Omega|$ into all possible one particle states
with comparable probabilities. This leads to a strong divergence:
the vacuum state disappears immediately. Because of positivity of
all corresponding quantities this divergence cannot be
renormalized. The difficulty remains when one goes to more
complicated theories because according to axiomatic field theory
no local operators can annihilate the vacuum state if all
conventional requirements are fulfilled. But without the positive
energy condition this result cannot be proved. That's why there
exist local tachyonic fields annihilating the vacuum state. Such
fields can be substituted into equation \eqref{TSsimple} for the
$\phi(x)$ without creating
divergences\footnote{To use tachyonic fields in this context
has proposed P.~R.~Pearle \cite{6}}.
The tachyonic field makes
the vacuum unstable by conventional tool, but if this field
interacts with other fields very weakly such instability may be
unobservable.

Let us look for a spin zero field $\phi(x)$ satisfying the
causality conditions
\begin{align*}
[\phi(x), \phi(y)] &= 0 \quad \forall\;  x, y : (x-y)^2 < 0, \\
[\phi(x), \phi^{\dagger}(y)] &= 0 \quad \forall\;  x, y : (x-y)^2 < 0
\end{align*}
and annihilating the vacuum state:
\begin{equation} \label{phikillsomega}
\phi(x) | \Omega \rangle = 0.
\end{equation}
Let us put
\begin{align}
F(x) &= i \langle \Omega | [\phi(x), \phi^{\dagger}(0)] | \Omega \rangle, \label{F}\\
F_r(x) &= \theta(x^0)F(x), \notag \\
F_a(x) &= \theta(-x^0)F(x), \notag
\end{align}
where $\theta(x^0)=1$ if $x^0>0$, $\theta(x^0)=0$ if $x^0<0$, and
define the Green function

\[
G(x) = i \langle \Omega | T(\phi(x)\phi^{\dagger}(0)) | \Omega \rangle,
\]
Due to the equality \eqref{phikillsomega} one gets $G(x)=F_r(x)$. Let us define
Fourier transform of the $F(x)$:

\begin{equation}\label{fourie}
\tilde F(k) = \frac{1}{(2\pi)^2}\int d^4xe^{ikx}F(x)
\end{equation}
and similarly $\tilde F_r(k) = \tilde G(k)$ and $\tilde F_a(k)$.

In analogy with conventional derivation of Lehmann representation one
gets

\begin{align}
\tilde G(k) &= \tilde F_r(k) = f(k^2 + i\epsilon \sgn k^0), \notag \\
\tilde F_a(k) &= f^* (k^2 + i\epsilon \sgn k^0), \notag \\
\tilde F(k) &= \tilde F_r(k) - \tilde F_a (k) = 2i \Im f(k^2 + i\epsilon \sgn k^0), \label{tildaF}
\end{align}
where the function $f(s)$ is analytic in all the complex $\mathbb
C$-plane without, perhaps, the positive part of real axis. From
 the  eq-ns \eqref{F} and \eqref{fourie} it follows that

\[
\Re \tilde F(k) = 0, \qquad \Im \tilde F(k) \ge 0,
\]
so that for each real $s \equiv k^2$
\[
\Im f(s \pm i\epsilon) \ge 0.
\]

To get the general representation for the $f(s)$ let us put

\begin{gather*}
\xi_+(s) = \frac12 (f(s) + f^*(s^*)), \\
\xi_-(s) = -\frac{i}{2\sqrt{-s}} (f(s) - f^*(s^*)),
\end{gather*}
where the $\sqrt{-s}$ is positive if $s<0$ and has a cut at $0 \le s <
+ \infty$. The functions $\xi_{\pm}(s)$ have the same analytic
properties as $f(s)$. Let us assume that the functions
$\xi_{\pm}(s)$ decrease at $|s| \rightarrow \infty$ faster than
$|s|^{-\eta}$ with some positive $\eta$. Then the dispersion
relations hold

\[
\xi_{\pm} = \frac{1}{\pi} \int\limits_{s'=0}^{\infty} ds' \frac{\Im \xi_{\pm}(s' + i\epsilon)}{s'-s}.
\]
Consequently it takes place the relation

\begin{equation}\label{f}
\begin{split}
f(s) = \frac{1}{2\pi} &\left( \int\limits_0^{\infty} ds' \frac{\Im f(s'+i\epsilon) - \Im f(s'-i\epsilon)}{s'-s} + \right.\\
&+ \left. i\sqrt{-s} \int\limits_0^{\infty} \frac{ds'}{\sqrt{-s'}} \frac{\Im f(s'+i\epsilon) + \Im f(s'-i\epsilon)}{s'-s} \right).
\end{split}
\end{equation}

Defining at $0 \le s \le + \infty$ arbitrary  nonnegative
decreasing fast enough functions $\Im f(s'+i\epsilon)$ and $\Im
f(s' -i\epsilon)$, one gets from the eq-n \eqref{f} the function $f(s)$
fulfilling all requirements.

One sees from the eq-ns \eqref{tildaF} and \eqref{f} that the function
$\tilde F(k)$  differs from zero at all $k^2 <0$ if it is not
zero everywhere. That's why by definitions
\eqref{F} and \eqref{fourie} the tachyonic spectrum is
continuous and contain all the negative part of the $k^2$-axis.

We can now put
\[
\phi(x) = \frac{1}{(2\pi)^2} \int d^4k\, a(k) e^{-ikx},
\]
where the $a(k)$ are annihilation operators fulfilling the relation

\begin{align*}
[a(k),a^{\dagger}(k)] &= \eta(k)\delta^4(k-k'), \\
[a(k),a(k')] &= 0, \\
a(k)|\Omega\rangle &= 0,
\end{align*}
where $\eta(k) = 2(2\pi)^2 \Im f(k^2 + i\epsilon \sgn k^0)$.
Inserting these operators $\phi(x)$ into the equation \eqref{TSsimple}
one sees that no divergence appears because the right side vanishes
if the $\rho$ is a vacuum state. All other purely tachyonic states
transform into the vacuum when the time goes.

If such a tachyonic fields interact with all other
nongravitational fields very weakly and with the gravitation in a
conventional way, they can exist without a contradiction with the
experiment. They can even be a part of the hidden matter in the
Universe. It is worth to investigate the possibility of existing
of such fields in cosmology.

But we have to take into account that the operator $\phi(x)$ of a
tachyonic field is not a macroscopic operator. That's why it is
not guaranteed that the equation of the type \eqref{TS} with such
a field will lead to the required decay of macroscopically
indefinite states. This point is not investigated today and it is
worth to do this.

Until now we have tried to describe the decay of macroscopically
indefinite states only formally. But one can ask what from the
additional terms in the Lindblad's equation can be derived. Much work has
been done to get such terms from the interaction of the object
with the environment (the thermal bath). The considerations are like
our simple example on the page \pageref{decoherence}. But the second
system is now the environment. Such considerations permit to get
quantitative results. Especially, the time of the decay of
macroscopically indefinite states can be calculated. In this way
it was shown that the macroscopically definite states are stable
against decay and the macroscopically indefinite are not, what is
an interesting result. It is to stress here that the needed
irreversibility arises not because the environment acts on the
system, but because some information about the system disappears
in the environment.

However these considerations do not solve the problem completely,
because the environment can be included into the quantum system (at
least in principle). So one has to consider a new larger
environment and so on. Finally all the Universe is included into
the quantum system, and  there is no environment more. One may
try to solve the problem asking oneself how some information about
the quantum system can disappear completely, i.e. in such a way
that it cannot return to the system. One may imagine three ways
leading to this disappearance.

Firstly the system may radiate electromagnetic, gravitational and
other waves that disappear at the infinity. This is possible in flat
as well as in curved space-time. Secondly in rapidly expanding
Universe the connection between two parts of a system may become
impossible because the appropriate light cones do not intersect.
Thirdly if our space-time is a surface in a space of higher
dimensions, the information can flow into the additional
dimensions.

At least in the first and second case a very long time is needed
before the information disappears completely. To fulfill the
principle of macroscopic definiteness one has to require that the
macroscopically indefinite states decay much more rapidly. So one
is led to consider a possibility that the future circumstances
influence the present state of affairs leading to the decay of
such states. This is in accordance with the Bell's proof that the
subquantum world cannot be causal in conventional relativistic
sense. If the only result of the action of the future on the present
is the decay of macroscopically indefinite states then this action
cannot contradict the conventional macroscopic causality.

Clearly this is only a very preliminary discussion of the problem.
But one has to bear it in mind.

\vskip 1em
{\bf Acknowledgments.}
The author is grateful to the UNESCO Regional Bureau
for Science in Europe (ROSTE) for supporting the International
V.~A.~Fock School for Advances in Physics (IFSAP).
This work was supported also by the Russian
Federation Ministry of Education, Grant No.~RNP.2.1.1.1112.


\begin{thebibliography}{1}
\newcommand{\enquote}[1]{``#1''}
\providecommand{\url}[1]{\texttt{#1}}
\providecommand{\urlprefix}{URL }
\expandafter\ifx\csname urlstyle\endcsname\relax
  \providecommand{\doi}[1]{doi:\discretionary{}{}{}#1}\else
  \providecommand{\doi}{doi:\discretionary{}{}{}\begingroup
  \urlstyle{rm}\Url}\fi
\providecommand{\eprint}[1]{\href{http://arxiv.org/abs/#1}{\texttt{#1}}}

\bibitem{1}
J.~S. Bell, \enquote{On the Einstein Podolsky Rosen paradox},
  \href{http://dx.doi.org/10.1103/PhysicsPhysiqueFizika.1.195}{\emph{Physics
  Physique Fizika}}, \textbf{1} (1964), 195--200.

\bibitem{2}
E.~C.~G. Sudarshan, P.~M. Mathews, J.~Rau, \enquote{Stochastic Dynamics of
  Quantum-Mechanical Systems},
  \href{http://dx.doi.org/10.1103/PhysRev.121.920}{\emph{Phys. Rev.}},
  \textbf{121} (1961), 920--924.

\bibitem{3}
V.~A. Franke, \enquote{On the general form of the dynamical transformation of
  density matrices}, \href{http://dx.doi.org/10.1007/BF01051230}{\emph{Theor.
  Math. Phys.}}, \textbf{27}: 2 (1976), 406--413.

\bibitem{4}
G.~Lindblad, \enquote{On the generators of quantum dynamical semigroups},
  \href{http://dx.doi.org/10.1007/BF01608499}{\emph{Communications in
  Mathematical Physics}}, \textbf{48} (1976), 119--130.

\bibitem{5}
A.~Bassi, G.~Ghirardi, \enquote{Dynamical reduction models},
  \href{http://dx.doi.org/10.1016/s0370-1573(03)00103-0}{\emph{Physics
  Reports}}, \textbf{379}: 5-6 (2003), 257--426,
  \eprint{arXiv:quant-ph/0302164}.

\bibitem{6}
P.~Pearle, \enquote{Relativistic collapse model with tachyonic features},
  \href{http://dx.doi.org/10.1103/PhysRevA.59.80}{\emph{Phys. Rev. A}},
  \textbf{59} (1999), 80--101, \eprint{arXiv:quant-ph/9902046}.

\end{thebibliography}

\end{document}